\documentclass[5p]{elsarticle}
\usepackage{hyperref}

\journal{Journal of Solid-State Electronics}

\begin{document}

\begin{frontmatter}

\title{Monolithic TCAD Simulation of Phase-Change Memory (PCM/PRAM) + Ovonic Threshold Switch (OTS) Selector Device}

\author[1]{M. Thesberg}
\author[1]{Z. Stanojevic}
\author[1]{O. Baumgartner}
\author[1]{C. Kernstock}
\author[2]{D. Leonelli}
\author[2]{M. Barci}
\author[3]{X. Wang}
\author[3]{X. Zhou}
\author[3]{H. Jiao}
\author[4]{G. L. Donadio}
\author[4]{D. Garbin}
\author[4]{T. Witters}
\author[4]{S. Kundu}
\author[4]{H. Hody}
\author[4]{R. Delhougne}
\author[4]{G. Kar}
\author[1]{M. Karner}

\affiliation[1]{organization={Global TCAD Solutions},
city={Vienna},
country={Austria}
}

\affiliation[2]{organization={Huawei Technologies R\&D Belgium N.V.},
city={Leuven},
country={Belgium}
}

\affiliation[3]{organization={HiSilicon Technologies},
city={Shenzhen},
country={China}
}

\affiliation[4]{organization={imec},
city={Leuven},
country={Belgium}
}

\begin{abstract}
Owing to the increasing interest in the commercialization of phase-change memory (PCM) devices, a number of TCAD models have been developed for their simulation. These models formulate the melting, amorphization and crystallization of phase-change materials as well as their extreme conductivity dependence on both electric field and temperature into a set of self-consistently-solved thermoelectric and phase-field partial-differential equations. However, demonstrations of the ability of such models to match actual experimental results are rare. In addition, such PCM devices also require a so-called selector device -  such as an Ovonic Threshold Switching (OTS) device - in series for proper memory operation. However, monolithic simulation of both the PCM and OTS selector device in a single simulation is largely absent from the literature, despite its potential value for material- and design-space explorations. It is the goal of this work to first characterize a PCM device in isolation against experimental data, then to demonstrate the qualitative behavior of a simulated OTS device in isolation and finally to perform a single monolithic simulation of the PCM + OTS device within the confines of a commercially available TCAD solver: GTS Framework.
\end{abstract}


\end{frontmatter}


\section{Introduction}

Due to the growing industry interest in phase-change memory (PCM) devices for commercial applications, a number of TCAD models have been developed for their simulation.\cite{scoggin2020field,woods2017modeling,ciocchini2014modeling,burr2012observation} These models formulate the melting, amorphization and crystallization as well as their extreme conductivity dependence on both electric field and temperature into a set of self-consistently-solved thermoelectric and phase-field models. However, any real PCM device must also have a selector device in series to stop the flow and heating of non-addressed cells in a memory array. Many device types have been suggested to play this role such as conventional devices like PN diodes and MOSFETs but also an unconventional Ovonic Threshold Switching (OTS) device\cite{burr2014access} has been suggested, which exploits the unusual \emph{electric} properties - rather than the melting properties - of amorphous phase-change materials in order to function as a selector diode.\cite{chien2018study} This is because phase-change materials have been observed to undergo a kind of reversible voltage breakdown above a critically-high electric field which allows for a kind of “turn on” behavior.\cite{ovshinsky1968reversible} This is often functionally described as a conductivity with separate thermally-dependent and electric field-dependent contributions that is dominated by the temperature-dependent portion at low fields, but owing to an exponential dependence on the applied electric field, becomes highly conductive in high-fields where the other term dominates (i.e. the conductivity, $\sigma$, behaves like $\sigma \propto \sigma_{T}(T) +  \sigma_{|E|}\exp(|E|)$ where $E$ is the electric field strength and $T$ is the temperature). Thus, in addition to the memory portion of the device, the OTS selector device also experiences important thermoelectric physics and heating that affects the memory portion as well as plays a role in design considerations like the minimization of thermal contamination (i.e. thermal cross-talk) between cells. And yet, the combined consideration of both memory and such selector devices is rarely considered despite its value for material and design space explorations. Here the goal is to demonstrate a combined simulation of both parts in a single monolithic simulation framework and show that all the important physics can be captured within one simulation.

Both the memory portion and selector portion of a PCM+OTS cell are ultimately made from phase-change materials, albeit those with different desired properties. The PCM part should be chosen to have ultra-fast crystallization, since crystallization-rate is the material property that most limits the maximum switching speed of such memories, where conversely the OTS portion should have a strong \emph{suppression} of crystallization. This is because it must remain amorphous during switching and operation as only in the amorphous phase where heating is limited (as Joule heating is low because the electrical current is low given the high resistivity of the amorphous state) and fields can get high is this  electrical “turn on” behavior observed. Thus both regions can be simulated with the basic models but modified with different material parameters for each region to exhibit either phase-changing or ovonic thresholding behavior as needed. The model used in this simulation is based off of \cite{woods2017modeling}, which will be discussed further in the next section, and was implemented into the GTS Framework,\cite{globalTcadSolutions} a commercial TCAD solver where the relevant partial-differential equations are solved within a finite-volume discretization scheme. This PCM+OTS study then proceeded in three steps. First, the behavior of the PCM portion alone is quantitatively validated against experimental data provided by imec in order to parameterize the model of \cite{woods2017modeling}. Second, the OTS portion alone is qualitatively demonstrated to exhibit the desired thresholding behavior. Finally, the combined PCM+OTS is considered.

\section{PCM-Only Device} 

\begin{table*}[t!]
\begin{center}
\begin{tabular}{|c|c|c|c|c|c|} 
 \hline
& $\sigma_0$ (S/m) & $c_{amorph}$ & $c_{nuc}$ & $c_{in-mesh}$ & $c_{inter-mesh}$  \\ \hline
2D & 35800 & 7.59$\times 10^{-3}$ & 7.33$\times 10^{-4}$ & 1.70$\times 10^{-9}$ & 2.51$\times 10^{-8}$\\ 
3D & 6050 & 7.01$\times 10^{-3}$ & 1.81$\times 10^{-3}$ & 1.46$\times 10^{-9}$ & 1.53$\times 10^{-8}$\\ 
 \hline
\end{tabular}
\caption{The optimal ``scaling coefficients'' obtained from matching experimental data for 2D and 3D simulation.\label{theTable}}
\end{center}
\end{table*}


The model used for simulation is that found in \cite{woods2017modeling}, which consists of models of the temperature- and field-dependence of the conductivity and Seebeck coefficient of phase-change materials based on look-up tables and non-linear functional dependencies as well as separate models for nucleation, domain growth and crystallization dynamics and a classical heating model which includes the Seebeck, Thompson, Peltier and Joule-heating effects. The complete model will not be recounted here (see \cite{woods2017modeling} for more details) but possesses upwards of a dozen possible variable parameters that can be tuned to match an experimental data set. However, in order to simplify the number of settable parameters instead here only five parameters were allowed to vary from the default values given in that work, where the stated parameters were not the result of a validation against experimental data but instead fine-tuned to produce plausible pulse behavior. These parameters were ``scaling coefficients'' of the nucleation, ``in-mesh'' growth, ``inter-mesh'' growth, amorphization rate and conductivity of Ge$_2$Sb$_2$Te$_5$ (GST). The meaning of a ``scaling coefficient'' is that, for example, the bulk conductivity of GST found in experiment was noticeably more resistive than literature values, however, there was not enough experimental data to fully chart the complete temperature and crystallization-fraction dependence of this resistivity. Thus, instead the look-up tables in \cite{woods2017modeling} were used with all the values being simply linearly-scaled to be more resistive by a single number. The scaling coefficients of the crystallization/amorphization rates have similar meanings. 

\begin{figure}[h!]
\centering
\includegraphics[width=0.45\textwidth]{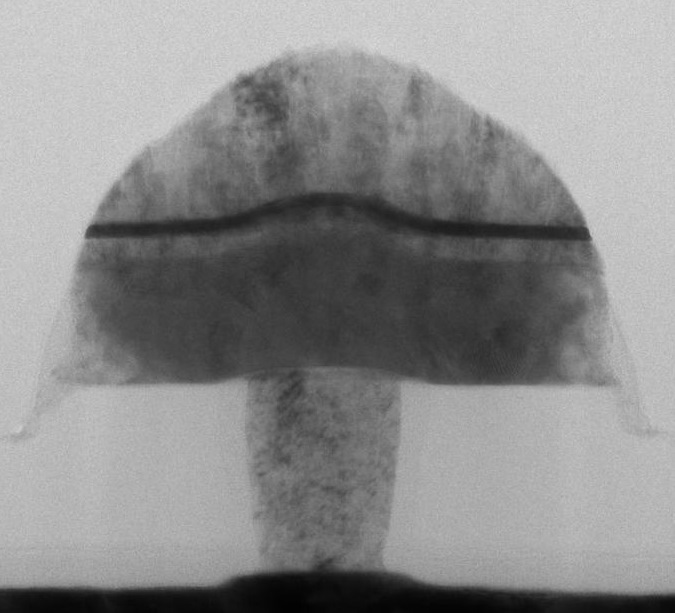}
\caption{TEM image of a PCM-only (i.e. no selector device is present) mushroom device with a 200 nm wide Ge$_2$Sb$_2$Te$_5$ (GST) PCM slab of 50 nm thickness contacted TiN cylindrical heater of 65 nm diameter.\label{fig:one}}
\end{figure}

\begin{figure}[h!]
\centering
\includegraphics[width=0.45\textwidth]{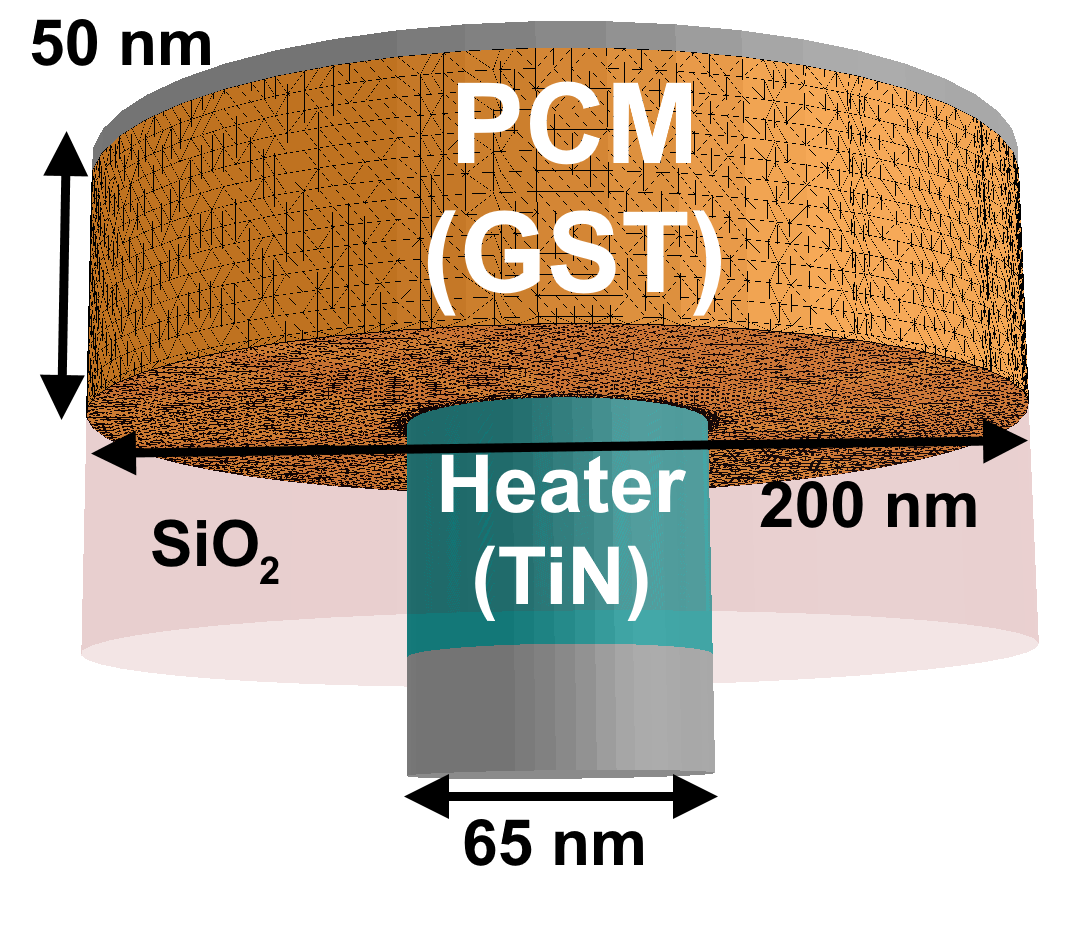}
\caption{Device schematic of PCM-only TCAD device used for parameterization and validation of phase-change properties against experimental data. This device slightly differs from the experimental device in that the PCM region is assumed a cylinder rather than a long rectangular prism in the out-of-plane direction like the original experimental device. However, such edge effects were found to have negligible contribution.\label{fig:two}}
\end{figure}

\begin{figure}[h]
\centering
\includegraphics[width=0.45\textwidth]{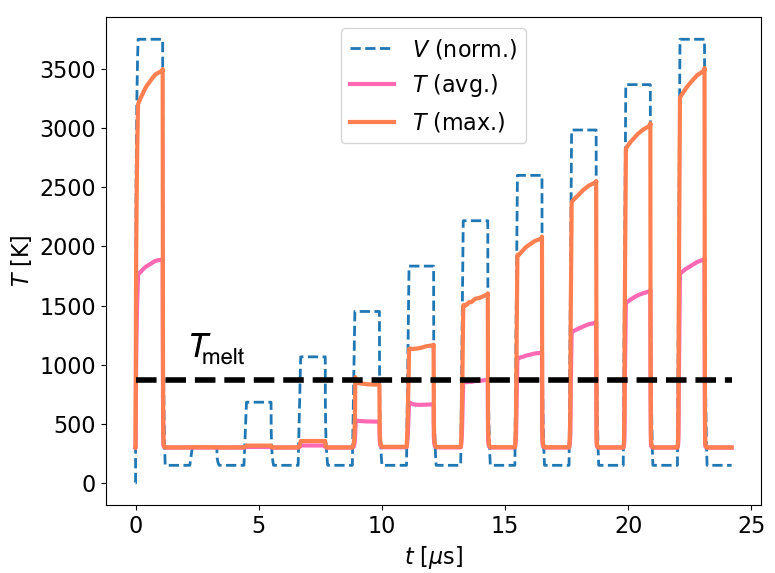}
\caption{The voltage (normalized to comparable units as the temperature, the highest pulse corresponds to 2.5 V) of the Reset-Set-Reset (RSR) experimental pulse  and resulting average and maximum temperature.  A 750 $\Omega$ resistor is also present in series with the device.\label{fig:three}}
\end{figure}

\begin{figure}[h]
\centering
\includegraphics[width=0.45\textwidth]{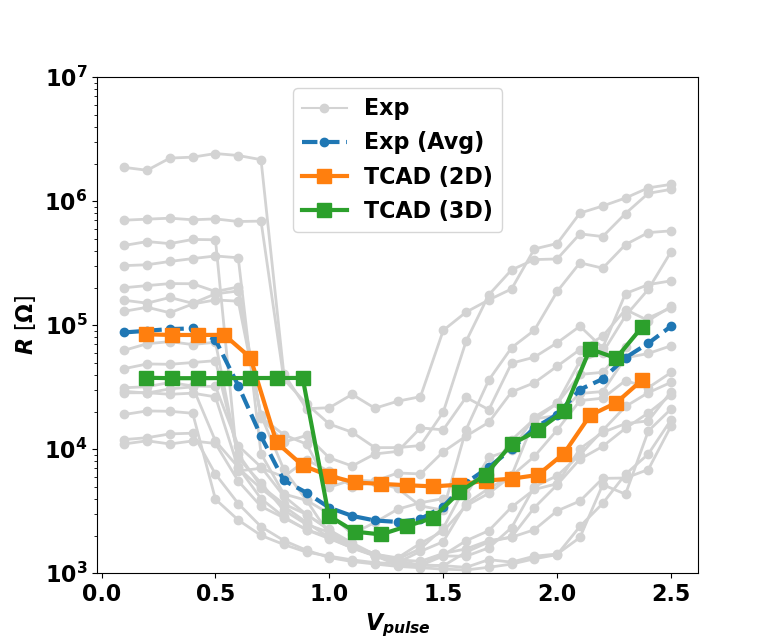}
\caption{Experimental resistance vs. applied pulse voltage data, along with the TCAD best-fit results in 2D and 3D, for the RSR pulse (750 $\Omega$ series resistor).\label{fig:four}}
\end{figure}

\begin{figure}[h]
\centering
\includegraphics[width=0.3\textwidth]{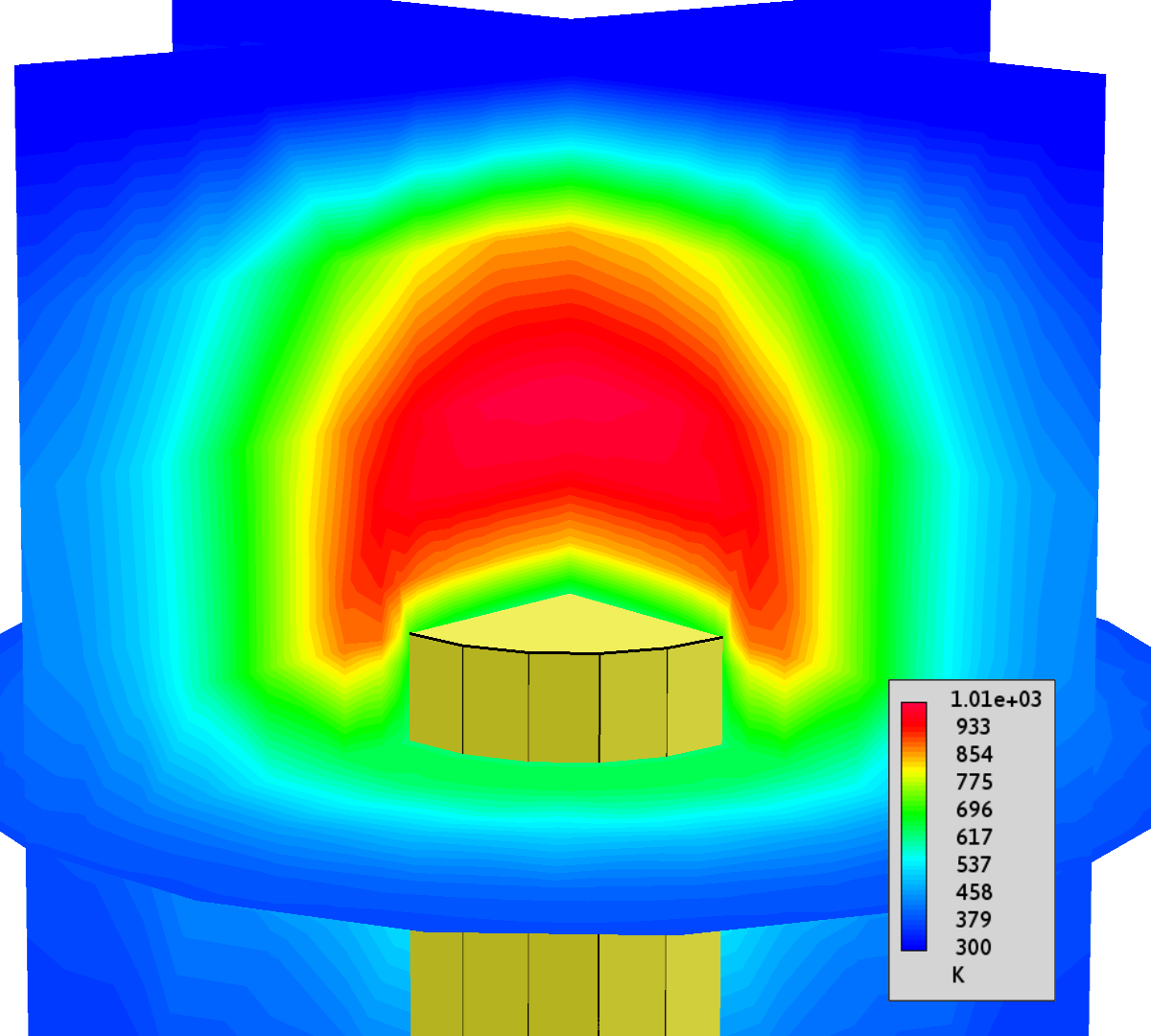}
\caption{Illustrative image of temperature distribution during melting in TCAD simulation.\label{fig:five}}
\end{figure}

In order to determine the value of these five parameters, experimental data was extracted from a mushroom cell (Figure \ref{fig:one}) consisting of a physical vapor deposition (PVD) GST layer of 50 nm thickness with a bottom ``heater'' TiN layer of 65 nm diameter and a top TiN electrode. A corresponding TCAD device was then created (Figures \ref{fig:two} and \ref{fig:three}) though with one key difference: the experimental device was a large rectangular prism in the out-of-plane direction where the TCAD device has cylindrical symmetry in order to speed-up simulation, which is valuable considering the large 5-dimensional parameter space that must be searched. The difference between a rectangular and cylindrical PCM region was found to be negligible since the far boundaries play little role in the over-all heat dissipation. 17 of the experimental devices were constructed and were then placed in series with a 750 $\Omega$ resistor and run through a ``Reset''-``Set''-``Reset'' (RSR) pulse sequence. All 17 devices were made on the same wafer with the same process though a great variation in final characteristics was observed. This RSR sequence involves the application of an initial 2.5 V/200 ns square pulse intended to put the device into the ``Reset''/amorphous state followed by an Incremental Step Pulse Programming (ISPP) scheme (Figure \ref{fig:three}) of 1 $\mu$s square pulses of increasing voltage followed by a read pulse of 0.1 V.  The resulting resistance vs. pulse voltage data is shown in Figure \ref{fig:four} for all 17 devices as well as the final TCAD results following a material parameter optimization search. The criteria of this search was chosen to be the logarithmic ensemble average. Due to the great device-to-device variation, this is believed to be optimal rather than an alternative approach of, for example, choosing a single representative device for validation, which may constitute a form over ``over-training''. The material parameter search was done both using two-dimensional and three-dimensional simulation and both results are given. The final optimal model parameters used are given in Table \ref{theTable}. Good agreement between TCAD and experiment is found. A representative image of the distribution of heat during a 3D TCAD simulation is given in Figure \ref{fig:five}.

\section{OTS-Only Device}

\begin{figure}[h]
\centering
\includegraphics[width=0.5\textwidth]{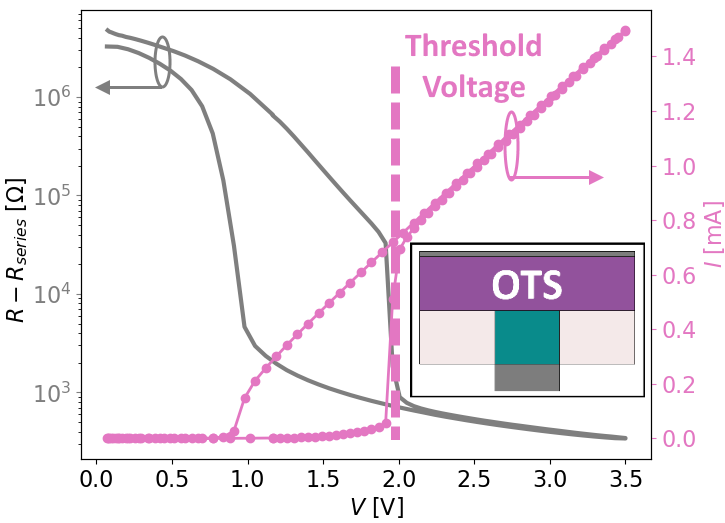}
\caption{Current and resistance vs. voltage behavior of the OTS-only device (with a 2 k$\Omega$ series resistance which has been subtracted off) showing a clear threshold voltage and ``turn on'' at $\sim$2.0 V for an applied 1 $\mu$s triangular pulse. The hysteresis observed does not reflect a phase transition of the OTS material - which remains amorphous throughout - but rather during the down-sweep the device is still at an elevated temperature and only returns to the low resistance state once cooled. \label{fig:six}}
\end{figure}

Owing to the current lack of experimental data for the OTS device at the time of publication, instead literature values for GST, including non-scaled conductivity and amorphization/crystal growth rates were used taken from \cite{woods2017modeling} -- which are different than those obtained from the parameterization of the PCM-only device as the obtained ``scaling coefficients'' are not used -- and modified such that the nucleation and so-called ``in-mesh'' and ``inter-mesh'' growth rates were set to zero. This ensures that the device always re-enters the amorphous state after melting. Thus, where the PCM device can be considered to behave as one that closely matches experiment, the OTS device should be considered as a ``representative'' ovonic switch. To verify that these model parameters do indeed exhibit the correct OTS behavior a simple 1 $\mu$s triangular pulse was simulated for a three-dimensional OTS structure identical in geometry to that used for the PCM-only simulation and the current versus voltage and well as resistance versus voltage results of which are shown in Figure \ref{fig:six}. In addition a higher series resistance of 2 k$\Omega$ was used rather than 750 $\Omega$ for the OTS and later PCM+OTS device in order to smooth out the numerics in the face of such dramatic exponential resistance changes. However, in Figure \ref{fig:six} the series resistance has been subtracted off.  A clear threshold turn on at a $V_{th}$ of $\sim$2.0 V is observed, verifying that the correct threshold switching behavior is being replicated.

\section{PCM+OTS Device}

\begin{figure}[h]
\centering
\includegraphics[width=0.3\textwidth]{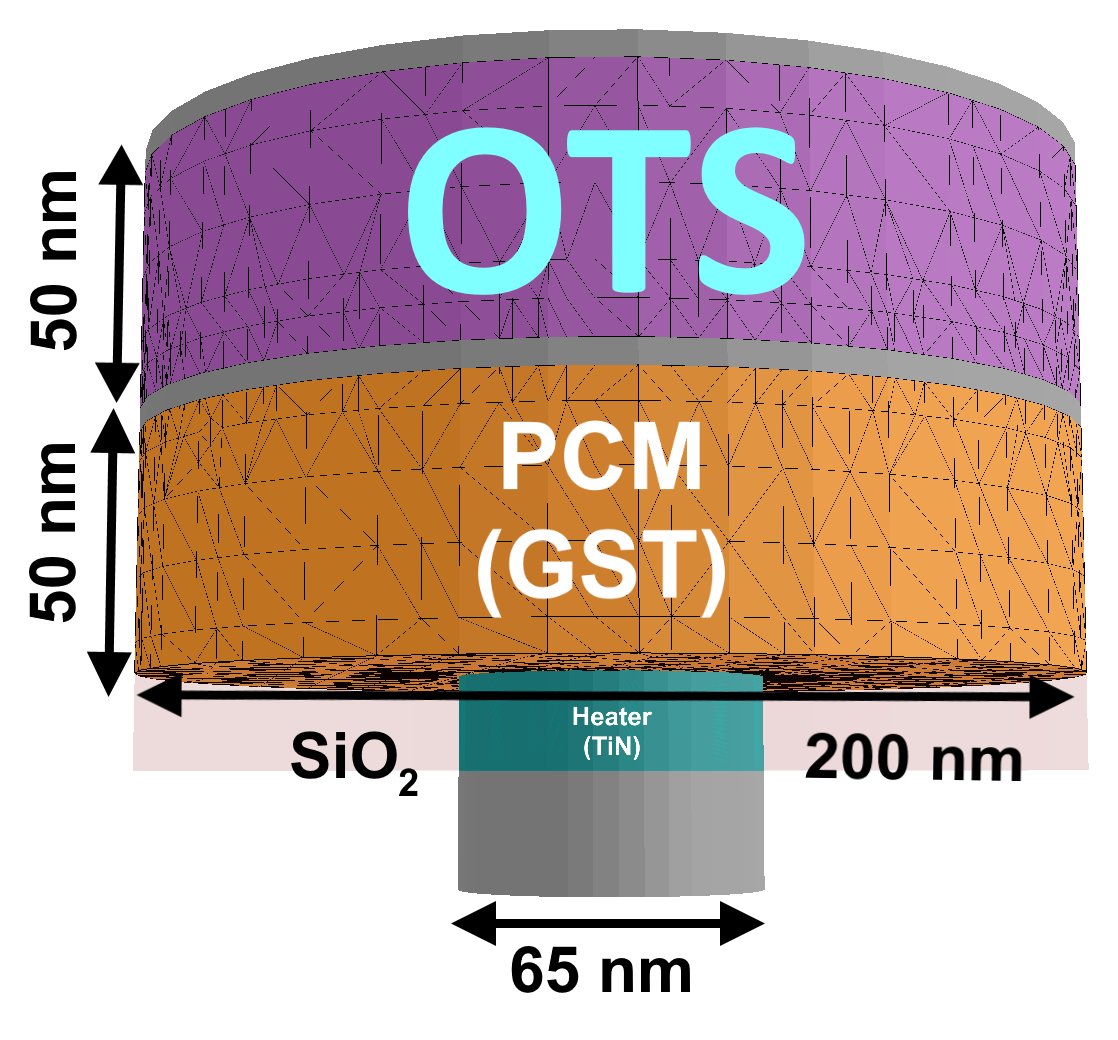}
\caption{Device schematic of the combined PCM+OTS device.\label{fig:seven}}
\end{figure}

\begin{figure}[h]
\centering
\includegraphics[width=0.5\textwidth]{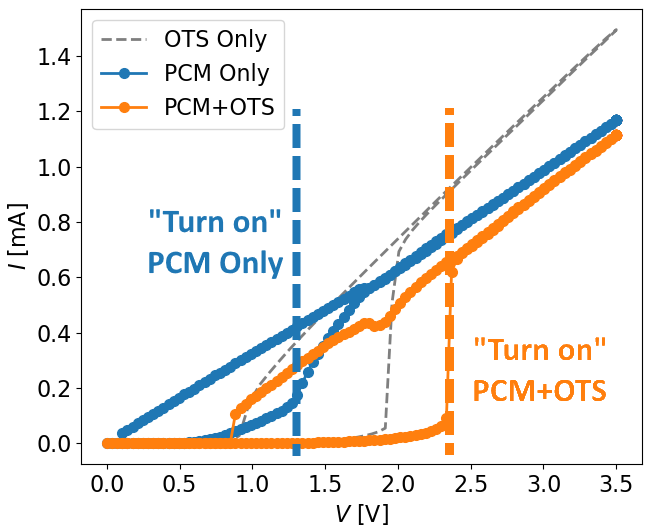}
\caption{Current vs. voltage behavior for the PCM-only (blue), OTS-only (grey dotted) and PCM+OTS (orange) devices as they are taken through a 1 $\mu$s set-reset pulse.\label{fig:eight}}
\end{figure}

\begin{figure}[h]
\centering
\includegraphics[width=0.5\textwidth]{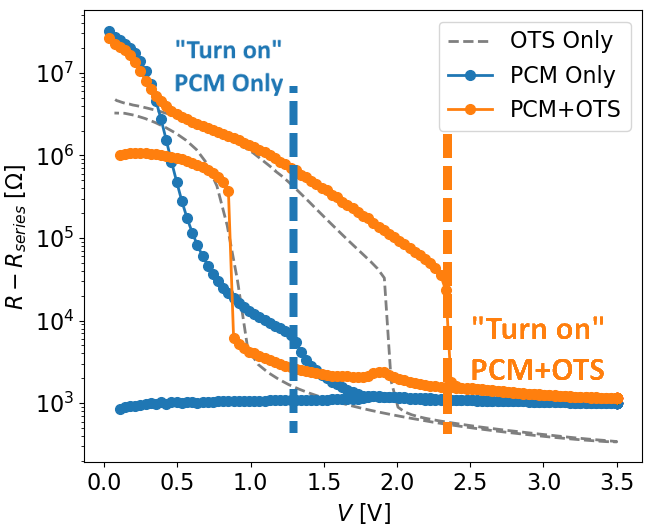}
\caption{Resistance (minus the 2 k$\Omega$ series resistance) vs. voltage behavior for the PCM-only (blue), OTS-only (grey dotted) and PCM+OTS (orange) devices as they are taken through a 1 $\mu$s set-reset pulse. The reason for the OTS-only device having a higher zero-voltage resistance than the PCM+OTS device is due to the different contacting situation in the OTS-only device (see text). \label{fig:nine}}
\end{figure}

Finally, the two materials with two different sets of parameters, separated by a thin TiN intermediate layer, were simulated as a monolithic device (Figure \ref{fig:seven}). As before the thicknesses of each layer was 50 nm and there is also a 2 k$\Omega$ series resistor present. This device was run through a 1 $\mu$s triangular pulse taking it from an initial ``Set'' state to a ``Reset'' state (Figure \ref{fig:eight}). For comparison the PCM-only device was put through the same pulse and the OTS-only results of Fig. \ref{fig:six} are provided for reference. In addition the resistance (minus the series resistance) is also plotted in Figure \ref{fig:nine}. The combined device behaves as expected with the OTS selector preventing flow at low voltages, preventing a non-addressed device from heating, while allowing flow at voltages above threshold. However, there are a few points of note in the combined device. Firstly, owing to the 50 nm thickness of the OTS device, the current of the PCM+OTS device is always lower than that of the PCM-only at all voltages due to the fact that the OTS layer provide a certain amount of series resistance. Furthermore, for an ideal PCM+OTS arrangement the threshold voltage of the OTS should be greater than half of the voltage required to flip the PCM+OTS device. This way a bit in a memory array can be flipped at the supply voltage while all other bits, held at half the supply voltage, draw negligible current since the OTS is below threshold. However, since the material parameters of the OTS device used here represent neither the results of material optimization nor have been quantitatively matched to experiment, any optimization of the design is outside the scope of this investigation where only the ability of monolithic simulation of both devices is being demonstrated. Finally, it can be noted in Fig. \ref{fig:nine} that the OTS-only device returns at zero voltage to a state of higher resistance than the PCM+OTS combined device even though the OTS layer in both devices is the same thickness. This is due to the fact that in the OTS-only device the OTS is sandwiched between a wide-area (i.e. 200 nm diameter) contact and a small-area TiN heater (65 nm diameter) where in the PCM+OTS device it is sandwiched between two wide-area conductive layers and thus its resistance in the two cases is not directly comparable.

\begin{figure}[h]
\centering
\includegraphics[width=0.49\textwidth]{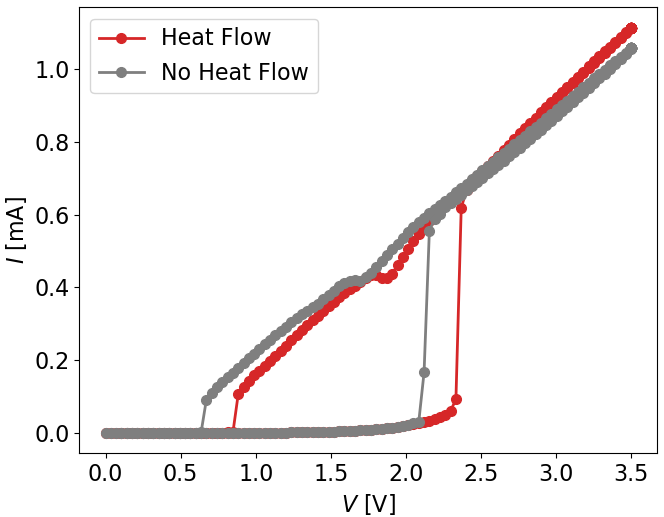}
\caption{Current vs. voltage behavior for otherwise identical PCM+OTS devices where the thermal conductivity of the intermediate layer is set to a realistic value for TiN or a negligible value (effectively preventing any thermal contamination). It can clearly be seen that ignoring thermal cross-talk produces noticeably different ``turn-on'' behavior. \label{fig:ten}}
\end{figure}

As a final consideration of the value of such monolithic simulation a demonstration of the clear effect of thermal contamination between the OTS and PCM devices, an effect that is missed when each are treated as separate circuit elements, is shown in Figure \ref{fig:ten}. Here the current vs. voltage characteristics of near-identical PCM+OTS structures is shown. The only difference between the structures is that in one the normal, high thermal conductivity of TiN is used for the intermediate layer separating the PCM and OTS layers and thus heat can freely move between the two in a realistic way and in the other the thermal conductivity of the middle layer is set to zero thus mimicking a case of ignoring the possibility of thermal contamination between the two. There is clearly a noticeable difference in ``turn-on'' behavior between the two cases since the OTS' low resistivity is a field-driven and not temperature-driven phenomena and thus for the heating of the PCM it acts merely as a detrimental heat-sink in the case where they share heat. Such effects demonstrate the importance of a monolithic approach.

Thus the ability to perform combined simulation of PCM+OTS selector in a single TCAD simulation has been demonstrated and can be used as a basis for further studies, such as investigation of coupling between the PCM and selector or of thermal contamination between neighbouring cells.


\end{document}